\documentclass[prd,floatfix,preprintnumbers]{revtex4}
\usepackage{amsmath}
\usepackage{graphicx,epsfig}
\usepackage{color}
\begin{document}
\title{How Slowly can the Early Universe Expand?} 
\author {Robert J. Scherrer}
\affiliation{Department of Physics and Astronomy, Vanderbilt University,
Nashville, TN  ~~37235}

\begin{abstract}
When the expansion of the universe is dominated by a perfect fluid with
equation of state parameter $w$ and a sound
speed $c_s$ satisfying $w = c_s^2 \le 1$, the Hubble parameter $H$ and time $t$
satisfy the bound $Ht \ge 1/3$.  There has been recent interest
in ``ultra-slow" expansion laws with $Ht < 1/3$
(sometimes described as ``fast expanding" models).  We examine various models
that can produce ultra-slow expansion: scalar fields with negative potentials,
barotropic fluids, braneworld models,
or a loitering phase
in the early universe.
Scalar field models and barotropic models for ultra-slow expansion are unstable
to evolution toward $w = 1$ or $w \rightarrow \infty$ in the former case
and $w \rightarrow \infty$ in the latter case.  Braneworld models
can yield ultra-slow expansion but require an expansion law beyond the
standard Friedman equation.  Loitering early universe models
can produce a quasi-static expansion phase in the early universe
but require an exotic negative-density component.  These results suggest
that appeals to an ultra-slow expansion phase in the early universe should
be approached with some caution, although the loitering early universe
may be worthy of further investigation.  These results do not
apply to
ultra-slow contracting models.
\end{abstract}

\maketitle

\section{Introduction}

Consider a universe dominated by a perfect fluid with equation of state
\begin{equation}
\label{w}
w = \frac{p}{\rho},
\end{equation}
where $p$ and $\rho$ are the pressure and density of the fluid,
respectively.
The evolution of $\rho$ as a function of the scale factor $a$ is given by
\begin{equation}
\label{rhoevol}
a \frac{d \rho}{da} = -3(\rho + p),
\end{equation}
so that
\begin{equation}
\rho \propto a^{-3(1+w)}.
\end{equation}
Since the Hubble parameter $H$ corresponds to
\begin{equation}
\label{H}
H \equiv \frac{\dot a}{a} = \sqrt{\frac{\rho}{3}},
\end{equation}
(where
we take $\hbar = c = 8 \pi G = 1$ throughout)
the scale factor $a$ evolves as a power of the time $t$:
\begin{equation}
\label{a(t)}
a \propto t^{2/3(1+w)},
\end{equation}
for $w > -1$.
In the standard cosmological model, the universe undergoes a period of radiation
domination ($w = 1/3$), followed by an epoch of matter domination ($w = 0$),
and is currently entering a period of dark energy domination with $w$ close to
$-1$.

However, many papers have considered the possibility of more exotic
evolution in the early universe.  For instance, the possibility that the
universe underwent a period in which the dominant fluid had a stiff equation
of state with $w=1$ and $a \propto t^{1/3}$ has been investigated
in relation to baryogenesis \cite{Joyce}, Big
Bang nucleosynthesis (BBN) \cite{stiffBBN}, the relic
abundance of dark matter \cite{Salati,VG,Erickcek,Deramo}, and the propagation of gravitational radiation
(\cite{Gouttenoire} and references therein).

More recently, there has been
speculation regarding the possibility of ultra-slow expansion.
For example, if the universe were dominated by a fluid with $\rho \propto a^{-n}$, and $n > 6$, then
$a \propto t^\alpha$ with $\alpha < 1/3$.  Since $a(t)$ is strongly
constrained from the era of BBN onward, most of the discusion of
these models concerns the calculation of relic particle abundances
\cite{Deramo,Bernal,Barman,Ganguly}, baryogenesis \cite{Chakraborty}, and leptogenesis \cite{Mahanta}.
Here we examine the underlying assumption of these papers:  are there
plausible models for which $Ht < 1/3$ over some range in $t$?  
Note that some of the terminology in the literature is confusing: an epoch
with $\rho \propto a^{-n}$, $n > 6$ is often referred to as ``fast expansion".
Such an epoch does correspond to fast expansion in the sense that $H(T)$ (where $T$ is the background
radiation temperature) is larger than in the standard model, with the ratio between $H$ in this nonstandard
model to $H$ in the standard model increasing with $T$.  However, the value of $H$ expressed
a function of time is smaller in the models we consider here than in the standard model, so we shall describe
these models as ultra-slow expansion.

Models with ultra-slow expansion are constrained by the fact
that the speed of sound, $c_s^2 = dp/d\rho$, is required to be sub-luminal,
so that $dp/d\rho \le 1$.  For a perfect fluid with $w = p/\rho$,
we have $w \le 1$ and $Ht < 1/3$.
There are, however, many models for which $dp/d\rho \ne p/\rho$, such
as scalar field models, barotropic fluids, and mixtures of perfect fluids,
all of which we discuss below to determine if they can provide a period
of expansion with $Ht < 1/3$.
We examine
previously-discussed models (scalar fields and braneworld expansion laws)
in Secs. 2 and 4, respectively, and models that have not been previously
explored (general barotropic fluids and loitering early universe models) in
Secs. 3 and 5.  Our results are summarized in Sec. 6.

\section {Scalar fields with negative potentials}

Scalar fields providing a significant component
of the energy density of the universe have frequently been invoked in cosmology.  They were first
introduced as the
main component of models for inflation (see, e.g., Refs. \cite{Lyth,Allahverdi} for reviews). Later, under the name ``quintessence," scalar
fields were investigated as an alternative to the cosmological constant as a mechanism to drive
the observed accelerated expansion
of the universe
\cite{RatraPeebles,Wetterich1,Wetterich2,Ferreira1,Ferreira2,CLW,CaldwellDaveSteinhardt,LiddleScherrer,SteinhardtWangZlatev}.
(See Refs. \cite{Copeland1,Bamba} for reviews).

The equation governing the evolution of a scalar field $\phi$ in a potential $V(\phi)$ is
\begin{equation}
\label{phievol}
\ddot{\phi} + 3H\dot{\phi} + \frac{dV}{d{\phi}} = 0,
\end{equation}
where the dot denotes the time derivative.
The pressure and density of the
scalar field are given by
\begin{equation}
\label{pphi}
p_\phi = \frac{\dot \phi^2}{2} - V(\phi),
\end{equation}
and
\begin{equation}
\label{rhophi}
\rho_\phi = \frac{\dot \phi^2}{2} + V(\phi),
\end{equation}
respectively, so that the equation of state parameter, $w$, from Eq. (\ref{w}) is
\begin{equation}
\label{wphi}
w = \frac{\dot\phi^2/2 - V(\phi)}{\dot \phi^2/2 + V(\phi)}.
\end{equation}
Note that the sound speed in these
models is $c_s^2 = 1$, independent of the value of $w$.

From Eq. (\ref{wphi}), it is clear that $-1 \le w \le 1$
as long as $V(\phi) \ge 0$.  However, negative potentials can give rise
to $w > 1$ and ultra-slow expansion or contraction.  Scalar field models
with negative potentials have long been of interest \cite{Felder,Heard,Macorra,Perivolaropoulos,Copeland2}.
In particular, it has been noted that a negative exponential potential of the form
\begin{equation}
\label{Vexp}
V(\phi) = - V_0 e^{-\lambda \phi},
\end{equation}
where $V_0 > 0$ and $\lambda > 0$, can lead to a constant
value of $w$ with $w \gg 1$ during the
contracting phase of the ekpyrotic model \cite{Khoury,Ijjas,Andrei}. [Note that the evolution of all of the physically-relevant parameters is invariant
if we take instead $V(\phi) = -V_0 e^{\lambda \phi}$ and invert the sign of $\dot \phi$].  It is then reasonable to
assume that this negative exponential potential can also yield ultra-slow expansion.  Indeed, D'Eramo et al.
\cite{Deramo}
provide a solution to the equation of motion with the potential of Eq. (\ref{Vexp}) that yields
just such an expansion:
\begin{equation}
\label{phitracker}
\phi = \phi_i + \frac{2}{\lambda}\ln\left(\frac{t}{t_i}\right),
\end{equation}
which gives an equation of state parameter
\begin{equation}
\label{wtracker}
w = \frac{\lambda^2}{3} - 1.
\end{equation}
Thus, for $\lambda > \sqrt{6}$, we have $w > 1$ and ultra-slow expansion.

The first thing to note about this solution is that it represents a scalar field rolling uphill in the potential,
which corresponds to very unnatural initial conditions.  Second, the solution is unstable.  This was first noted
by Heard and Wands \cite{Heard}, who investigated the evolution of scalar fields with both positive and negative
exponential potentials (see also the later discussion in Ref. \cite{Copeland2}).  The corresponding solution
for a contracting universe, on the other hand, is stable.

To illustrate the nature of this instability, and to investigate the rate at which
small perturbations to the fixed-point solution grow,
consider the expression for the evolution of $w$ for a scalar field in an exponential potential
of the form given by Eq. (\ref{Vexp})
in the limit where the scalar field dominates the expansion \cite{Linderw,ScherrerSen}:
\begin{equation}
\label{wprime}
a\frac{dw}{da} = (w-1)\sqrt{3(1+w)}\left[\sqrt{3(1+w)} - \lambda \right].
\end{equation}
The derivation of Eq. (\ref{wprime}) assumes that $\dot \phi >0$, but it is
trivial to generalize it to the opposite case.
As expected, this equation has a solution of the form $dw/da = 0$ and $w$
given by Eq. (\ref{wtracker}), corresponding to the last factor on the right-hand side
of Eq. (\ref{wprime}) equal to zero.
Now suppose we perturb this solution with a small change to $w$.  For $w > 1$, a positive change in $w$
gives $dw /da > 0$ and $w \rightarrow \infty$, while a negative change yields $dw/da < 0$
and $w \rightarrow 1$.  Thus, the solution given by Eqs. (\ref{phitracker}) and (\ref{wtracker})
is unstable.
Note that the opposite is true for $w < 1$; in this case the solution is stable.

Of course, even a transient solution with $w >1$ could be sufficient to produce interesting changes
to the evolution of relic particle densities, as discussed in Refs.
\cite{Deramo,Bernal,Barman,Ganguly,Chakraborty,Mahanta},
so it is important to examine the rate at which $w$ evolves away from its unstable fixed-point value.
If we write $w = w_0 + \Delta$, with $w_0 = \lambda^2/3 -1$ and expand Eq. (\ref{wprime}) to linear order
in $\Delta$, we obtain
\begin{equation}
a \frac{d\Delta}{da} = \frac{3}{2}(w_0 -1)\Delta,
\end{equation}
so the evolution of $w$ near $w_0$ is
\begin{equation}
w = w_0 + \Delta_0 a^{(3/2)(w_0-1)},
\end{equation}
where we are defining $a=1$ to be the scale factor at which $w = w_0 + \Delta_0$.
Thus, $w$ diverges from its fixed-point value $w_0$ as a power of the scale factor, with this power increasing
for larger values of $w_0$.  As an example, the largest value of $w$ considered in Ref. \cite{Deramo} is
$w = 5/3$, for which the density of the scalar field scales as $\rho_\phi \propto a^{-8}$.  In this
case, we have $w = w_0 + \Delta_0 a$. The effects examined in Ref. \cite{Deramo} require
this value of $w$ to be maintained over a factor $\sim 100$ in scale factor.  In order to achieve
this, $w$ needs to be initially tuned to within less than $1 \%$ of its fixed-point value.

\section{Barotropic models}

As noted in the introduction, a perfect fluid with $w = p/\rho$ has a sound speed $c_s^2 = w$, so that
$c_s^2 \le 1$ forces $w \le 1$.  However, we can break the equivalence between $w$ and $c_s^2$ by going
to a more complex relation between $p$ and $\rho$.  In
barotropic models, the pressure is a fixed function of the density:
\begin{equation}
\label{baro}
p = f(\rho).
\end{equation}
(Perfect fluids are the special case for which $p = w\rho$).  Barotropic models
have been studied extensively as possible models for dark energy.
Particular models of this form include
the Chaplygin gas \cite{Kamenshchik,Bilic} and 
the generalized Chaplygin gas \cite{Bento}, the linear/affine equation of state 
\cite{linear1,linear2,AB,Quercellini}, the quadratic 
equation of state \cite{AB}, the Van der Waals equation of state
\cite{VDW1,VDW2}, and more complicated equations of state \cite{Nojiri,Stefancic}.
A general study of the properties of barotropic models for dark energy
was undertaken in Refs. \cite{LinderScherrer,Bielefeld}. Note
that there is a simple mapping between the barotropic models discussed here and
purely kinetic $k$-essence
models \cite{LinderScherrer}, so the results presented here can be extended in a straightforward way
to the latter set of models.

These models seem plausible as a source for ultra-slow expansion because we now have $dp/d\rho \ne p/\rho$,
so that one can construct models for which $c_s^2 = dp/d\rho \le 1$ but $p/\rho > 1$.  As an example,
consider one of the simplest barotropic models, in which $p$ is a linear function of $\rho$
\cite{linear1,linear2,AB,Quercellini}:
\begin{equation}
p = p_0 + \alpha \rho,
\end{equation}
where $p_0$ and $\alpha$ are constants.  The requirement that $c_s^2 \le 1$ gives $\alpha \le 1$.
Using Eq. (\ref{rhoevol}), we find the following relation between the density and scale factor:
\begin{equation}
\label{rholinear}
\rho = Ca^{-3(\alpha+1)} - \frac{p_0}{\alpha +1},
\end{equation}
where $C$ is a constant.  In the limit of early times (small $a$), the first term dominates and the
density simply scales as a perfect fluid with $w = \alpha$.  When the two terms on the right-hand
side of Eq. (\ref{rholinear}) become of roughly equal magnitude, there is a transient period for which
$w > 1$ and the expansion becomes ultra-slow.  However, we rapidly have $w \rightarrow \infty$ and
$\rho \rightarrow 0$.

This fate for barotropic fluids with $w > 1$ is not peculiar to this particular choice of model; it is
generic to all such models.  To see why this is the case, we use the expression for the evolution of $w$
in barotropic models given in Ref. \cite{LinderScherrer}:
\begin{equation}
a \frac{dw}{da} = 3(1+w)(w - c_s^2).
\end{equation}
If $c_s^2 \le 1$ (as required) and $w > 1$, then the evolution is
manifestly unstable; $dw/da > 0$ and $w$ rapidly evolves to $\infty$.
Hence, it does not appear that any barotropic model can provide sustained ultra-slow expansion.

\section{Braneworld models}

In braneworld cosmologies, the observable universe is a brane embedded in a higher-dimensional bulk.
The standard-model fields are confined to our 3-brane, while gravity alone propagates in the bulk.
The most widely-investigated model of this type is the type II Randall-Sundrum model, in which
the brane has positive tension and the bulk contains a positive cosmological constant \cite{RS1,RS2}.
In such models, in flat spacetime, we can write the Hubble parameter as
\begin{equation}
H^2 = \frac{\rho}{3}\left(1 + \frac{\rho}{\rho_0}\right) + \frac{C}{a^4} + \frac{\Lambda}{3},
\end{equation}
where $C$ and $\rho_0$ are constants, with the latter depending on the 5-dimensional Planck mass,
and $\Lambda$ is the cosmological constant.  At early times, if we neglect $C$, the Hubble parameter
evolves as $H \propto \rho$ instead of $H \propto \sqrt{\rho}$.  The effects of this altered
evolution on the relic dark matter abundance were investigated in Refs. \cite{Okada,Dahab,OkadaOkada,
Meehan}.
Ref. \cite{Dahab}, in particular, noted that this modification to the Friedman equation could lead
to ultra-slow expansion.  When $H \propto \rho$, we have, instead of Eq. (\ref{a(t)}),
\begin{equation}
a \propto t^{1/3(1+w)}.
\end{equation}
Thus, 
even a radiation-dominated universe leads to ultra-slow expansion at early times, with $a \propto t^{1/4}$.
For an arbitrary perfect fluid, with $w \le 1$, the slowest possible expansion is
given by $a \propto t^{1/6}$.

It is clear that braneworld models can produce evolution slower than the bound set by a stiff perfect fluid.  However,
such behavior comes with an expansion law that does not obey the standard Friedman equation, so that standard calculations
of, e.g., relic particle evolution must incorporate this modified expansion law, as was done in Refs. \cite{Okada,Dahab,OkadaOkada,Meehan}.

\section{A loitering early universe}

The loitering universe is an idea that goes back to Lemaitre \cite{Lemaitre}.  This model
requires a spatially closed universe containing matter and a cosmological constant, so that $H$ is given by
\begin{equation}
\label{loiter}
H^2 = \frac{\rho_{M0}}{3}\left(\frac{a}{a_0}\right)^{-3} + \frac{\Lambda}{3} - \frac{\kappa}{a^2},
\end{equation}
where $\rho_{M0}$ and $a_0$ are the present-day matter density and scale factor, respectively, $\Lambda$ is the (positive)
cosmological constant and $\kappa = 1$ is the curvature.  The last term in Eq. (\ref{loiter}) must be large enough to alter
the evolution of $H$ at late times, but not so large that $H$ ever reaches $0$, which corresponds to recollapse.  There has
been persistent interest in this model \cite{Sahni}, but it is clear that it does not correspond to our best current observations; in particular,
the curvature of the universe is now known to be very small \cite{Ade}.  To remedy this defect,
Sahni and Shtanov \cite{SahniShtanov} proposed a loitering model in the context of braneworld models.  This model allows loitering
to occur in a flat universe and at somewhat higher redshifts than in the original model described by Eq. (\ref{loiter}).

Neither model, however, can accommodate loitering in the early universe.  The reason is that both of these models evolve 
asymptotically to
a cosmological-constant dominated evolution.  This is desirable at the present, but cannot be incorporated into an acceptable evolution
in the early universe.  Instead, we require a loitering solution that evolves asymptotically to a radiation-dominated early universe
consistent with observations.  (Note also that Ref. \cite{Brandenberger} proposed a loitering phase in the
context of Brane Gas Cosmology; this phase
takes place in the very early universe, before the epoch of interest here).

In analogy to Eq. (\ref{loiter}) we seek a universe containing multiple perfect fluids, but which is dominated by radiation (with
density $\rho_\gamma \propto a^{-4}$) at late times.  Therefore, any additional fluids we add to the radiation
must have a density that decreases faster than $a^{-4}$.  The fastest possible
decay occurs for a stiff fluid with $c_s^2 =1$ and $\rho_S \propto a^{-6}$.  Finally,
to allow for a loitering phase, we add a component with an intermediate equation of state and negative energy density,
$\rho_5 \propto a^{-5}$ and $\rho_5 < 0$.  Then for this mixture of fluids, the total density is given by
\begin{equation}
\label{rholoiter}
\rho = \rho_{\gamma i}\left(\frac{a}{a_i}\right)^{-4} - \rho_{5 i} \left(\frac{a}{a_i}\right)^{-5} + 
\rho_{S i} \left(\frac{a}{a_i}\right)^{-6},
\end{equation}
where $a_i$ is an arbitrary fiducial value of the scale factor at which
the densities of the three components are given by
$\rho_{\gamma i}$, $\rho_{5 i}$ and $\rho_{S i}$.
The first and second derivatives of the scale factor are given by $(\dot a/a)^2 = (1/3) \rho$
and $\ddot a/a = -(1/6) (\rho + 3p)$.
Setting $\dot a = \ddot a = 0$ at $a=a_i$ gives
$\rho_{5 i} = 2 \rho_{\gamma i}$ and $\rho_{Si} = \rho_{\gamma i}$.  Choosing $\rho_{Si}$, $\rho_{5i}$
and $\rho_{\gamma i}$
near these values will then give loitering behavior for $a$ near $a_i$.  This is illustrated
in Fig. 1, where we set $\rho_{5 i} = 2 \rho_{\gamma i}$ and choose $\rho_{S i}$ close to $\rho_{\gamma i}$ to
produce varying degrees of loitering.
\begin{figure}[t]
\centerline{\epsfxsize=3.7truein\epsfbox{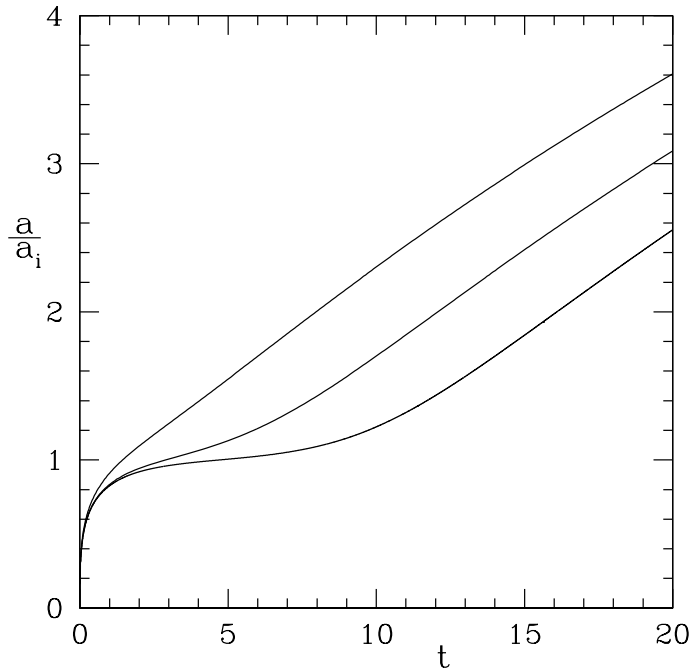}}
\caption{The evolution of the scale factor $a$ as a function of time $t$
for the loitering early universe produced by the density
given in Eq. (\ref{rholoiter}) with $\rho_{5 i} = 2 \rho_{\gamma i}$
and $\rho_{Si}/\rho_{\gamma i} = $ (top to bottom) $1.1, 1.01, 1.001$.}
\end{figure}

It is clear from Fig. 1 that the density given by Eq. (\ref{rholoiter}) can lead to a quasi-static
expansion or simply a significant slowing of the expansion rate, depending on the relative
densities of the components.  However, this model has several problems.  It requires a fine-tuning of the ratios
of the
densities in Eq. (\ref{rholoiter}), although this is also true of the original late-time loitering universe.
It also introduces a poorly motivated component with negative energy density, while in
the standard loitering model, the curvature automatically behaves as an effectively negative energy
density with the desired evolution.  Models with pairs of perfect fluids having, respectively,
positive and negative densities were examined in Ref. \cite{Nemiroff}, which also gives some motivation
for the latter.  Despite these problems, the evolution produced by this model is rather intriguing.
It is capable of producing an expansion rate slower than any of the other models considered here; for
an appropriate choice of parameters the universe can approach a nearly static state for a significant
period of time.

\section{Conclusions}
We find that there are no compelling models for ultra-slow expansion within the framework of the
standard Friedman equation.  Scalar field models and barotropic models are both unstable.  Braneworld
models do allow for ultra-slow expansion but only in the context of a modified expansion law.  The
loitering early universe introduced in the previous section is perhaps the most interesting, as it
allows for a nearly static phase in the early universe.  It would be interesting to explore such
a phase in connection with models for relic particle evolution and baryogenesis.  However, this model
requires a negative energy component with an unusual equation of state, and it is not clear that
there is a plausible motivation for such a component.

\end{document}